\documentclass[sigconf, authorversion, nonacm]{acmart}

\makeatletter
\def\@ACM@checkaffil{
    \if@ACM@instpresent\else
    \ClassWarningNoLine{\@classname}{No institution present for an affiliation}%
    \fi
    \if@ACM@citypresent\else
    \ClassWarningNoLine{\@classname}{No city present for an affiliation}%
    \fi
    \if@ACM@countrypresent\else
        \ClassWarningNoLine{\@classname}{No country present for an affiliation}%
    \fi
}
\makeatother

\usepackage{algorithm2e}
\usepackage{grffile} 
\usepackage{xcolor}
\usepackage{supertabular} 
\usepackage{cleveref}
\usepackage{multirow}
\usepackage{listings}
\usepackage{setspace}

\usepackage{newfloat}
\DeclareFloatingEnvironment[
   fileext=lol,
   name=Listing
]{listing}

\usepackage{natbib}

\usepackage{subcaption}
\DeclareCaptionSubType{listing}

\usepackage{mathtools} 
\newtheoremstyle{sltheorem}
{0pt}                
{0pt}             
{\slshape}        
{}                
{\bfseries}       
{.}               
{0.3em}               
{}                

\theoremstyle{sltheorem}

\AfterEndEnvironment{challenge}{\noindent\ignorespaces}

\AfterEndEnvironment{Capability}{\noindent\ignorespaces}

\newtheorem{Strategy}{Strategy}
\AfterEndEnvironment{Strategy}{\noindent\ignorespaces}

\usepackage{enumitem}
\usepackage{lipsum}
\setlist[itemize]{leftmargin=*}

\begin{document}

\title{The Agentic Regulator: Risks for AI in Finance and a Proposed Agent-based Framework for Governance}

\author{Eren Kurshan}
\affiliation{
 \institution{Princeton University}
 \city{Princeton}
 \state{New Jersey}
}
\email{ekurshan@princeton.edu}

\author{Tucker Balch}
\affiliation{
 \institution{Emory University}
 \city{Atlanta}
 \state{Georgia}
}
\email{tucker.balch@emory.edu}

\author{David Byrd}
\affiliation{
 \institution{Bowdoin College}
 \city{Brunswick}
 \state{Maine}
}
\email{d.byrd@bowdoin.edu}

\begin{abstract}
 Generative and agentic artificial intelligence is entering financial markets faster than existing governance can adapt. Current model-risk frameworks assume static, well-specified algorithms and one-time validations; large language models and multi-agent trading systems violate those assumptions by learning continuously, exchanging latent signals, and exhibiting emergent behavior. Drawing on complex adaptive systems theory, we model these technologies as decentralized ensembles whose risks propagate along multiple time-scales. We then propose a modular governance architecture. The framework decomposes oversight into four layers of “regulatory blocks”: (i) self-regulation modules embedded beside each model, (ii) firm-level governance blocks that aggregate local telemetry and enforce policy, (iii) regulator-hosted agents that monitor sector-wide indicators for collusive or destabilizing patterns, and (iv) independent audit blocks that supply third-party assurance. Eight design strategies enable the blocks to evolve as fast as the models they police. A case study on emergent spoofing in multi-agent trading shows how the layered controls quarantine harmful behavior in real time while preserving innovation. The architecture remains compatible with today’s model-risk rules yet closes critical observability and control gaps, providing a practical path toward resilient, adaptive AI governance in financial systems.

\end{abstract}


\settopmatter{printfolios=true} 
\maketitle

\section{Introduction \& Background}

AI adoption is accelerating across financial services \citep{Heaton16,Neurips19,BloombergGPT}, with sector investment projected to reach \$97 billion by 2027 \cite{WEF24}. Recent demonstrations show generative and agentic AI tackling increasingly complex use cases \cite{KPMG24,FinGPT}: large language models support financial analysis and advisory \cite{GPT4,GEMINI,Anthropic,Llama3,DeepSeek}, transformers aid time-series modeling and fraud detection \cite{TimeGPT}, multi-agent and agentic systems advance trading, risk management, portfolio optimization, and market simulation \cite{Xie,Park}, and AI agents are proposed for core model development tasks \cite{Okpala}.

These capabilities strain traditional risk-management and governance approaches \cite{AgenticReg,Schneider}. Even non-agentic systems exceed the practical limits of existing frameworks, while agentic AI amplifies risks related to safety, liability, and autonomy \cite{DeloitteAgent,ECOA,Weber2024,HDMA}. Despite active regulatory efforts, a comprehensive governance regime for AI in finance has not yet emerged \cite{EUAI,ChinaAI}.

\subsection{Motivation and Scope}

Prior to the emergence of generative AI, financial regulation faced an ``Innovation Trilemma'' \cite{Yadav}: only two of three key goals, regulatory clarity, market integrity, and innovation, could be pursued simultaneously. GenAI now compounds these tensions. Despite modernization efforts, traditional Model Risk Management (MRM) struggles to cope with the inherent complexity, unpredictability, and opacity of high-parameter models.

Generative AI systems are \textit{high-parameter complex systems}, marked by nonlinear interactions and emergent behavior that challenge regulatory design. AI-native governance approaches may offer a way forward, but they blur the line between system properties (e.g., trust, controllability) and regulatory mechanisms themselves. As a result, regulating these systems intersects directly with core AI safety challenges, including the alignment problem. Alarmingly, 63\% of financial firms have already deployed GenAI systems, and 35\% are piloting them \cite{GoogleCloud}, highlighting the urgency of governance innovation. This paper outlines key risks and proposes a multi-agent regulatory framework to address them.

\subsection{MAS Through a CAS Lens}

Multi-agent systems (MAS) and complex adaptive systems (CAS) both feature networks of interacting entities, but differ in purpose and origin. MAS stem from engineering: agents are programmed with goals, communication rules, and coordination algorithms to accomplish tasks \cite{balch1998behavior}. CAS, by contrast, emerge in nature and society, where global behavior arises from decentralized adaptation and local interactions \cite{holland1992complex}. MAS aim for engineered performance; CAS are studied to understand phenomena like self-organization and phase shifts.

MAS already power market simulations and agent-based prototypes, where learning agents compete and adapt \cite{byrd2020abides,byrd2022learning}. These dynamics enable stress-testing but also pose governance risks, as agents can learn manipulative or collusive strategies. 

From a CAS perspective, both MAS and GenAI systems exhibit emergent, high-dimensional, and adaptive behavior—properties that defy static MRM assumptions like fixed data, stable models, and one-time validations. 

In this work, we use MAS to design our regulatory architecture and CAS to analyze its behavior. The framework partitions oversight into interacting regulatory blocks: self-regulation components at the model level, firm-level governance modules, and external audit units. As a multi-agent system, these blocks adapt and coordinate while enforcing safety and compliance constraints. This modular design aligns governance architecture with the systems it must regulate—offering more flexible and resilient oversight than centralized, linear controls.

\section{The State of AI Regulation and Governance}

Adding to this complexity, a growing number of AI regulations are being enacted globally \cite{USAI, EUAI, ChinaAI, JapanAI, CanadaAI}, presenting significant complications for financial firms with global operations. The diversity of regulatory approaches highlights the need for novel AI governance systems and automation. Recent AI regulations follow fundamentally different regulatory philosophies and yield to dramatically different development requirements:

\textit{(i) Principles-Based Approach:} Provides general guidance without detailed rules or interpretations (e.g. Australia’s 2019 AI Framework \cite{AustraliaAI}). 

\textit{(ii) Risk-Categorization-Based Approach:} Classifies AI systems by their potential risk with stricter oversight for higher risk models (E.U. AI Act \cite{EUAI}). 

\textit{(iii) Rule, Process \& Standard-Based Approach:} Defines specific rules, standards, and procedural requirements (China's 2023 GenAI Interim Measures \cite{ChinaAI}). 

\textit{(iv) Result-Based Approach:} Enforces desired outcomes without prescribing detailed rules or processes (Singapore’s 2019 AI Governance Framework \cite{SingaporeAI}). 

This highlights the difficulty of satisfying all four regulatory requirements within a single model development and risk management framework.

Model risk management processes evolved primarily for quantitative mathematical models used in finance with guidance from regulations like SR 11-7 / OCC 2011-12 \cite{SR11-7}. In addition to missing critical AI regulatory needs like safety assurance, ethics, autonomy control, and emergent behavior management, MRM stages are ineffective for GenAI:

\textit{(i) Model Risk Rating}: Common static model and stable data assumptions have become obsolete in the latest generation of AI \cite{Abiyoke}. 

\textit{(ii) Initial Model Validation}: One-time validation of assumptions, code, and performance is inadequate to account for dynamic, non-deterministic behavior inherent in generative and agentic systems.

\textit{(iii) Ongoing Monitoring}: For generative and agentic AI, periodic reviews are insufficient; it needs real-time, adaptive oversight, making traditional monitoring obsolete. Fintechs do leverage more general continuous model monitoring, but traditional financial institutions are often still dependent on manual periodic reviews with KPI monitoring \cite{Abiyoke}.

Despite widespread efforts, progress in developing effective model governance and regulatory solutions for generative AI has been limited. Increasing model development costs and MRM challenges have driven financial firms towards a membership strategy from foundational AI model developers, but these increasingly centralized general-purpose AI models have limited support for financial regulations, safety or ethics, and thus pose growing risks.

Financial regulators are rapidly developing new frameworks and regulations for generative and autonomous AI. In the U.S., SEC, OCC, FINRA, and CFTC are all actively engaged in planning and discussions for AI, focusing on risks, governance, and supervisory controls. The Federal Reserve and FDIC have formed internal AI groups to develop guidance, reflecting a strong push for proactive and adaptive AI oversight in finance.

\section{Risks of AI in Financial Systems}

Existing MRM and governance frameworks lack the capacity to address the scale and severity of the emerging risks of AI, which requires a paradigm shift in model governance.  This section aims to cover a selection of emerging risks from GenAI (full coverage is beyond the scope of this paper). Generative AI exacerbates existing risks and introduces novel risks.  We enumerate them below.

\subsection{Systemic Risks of AI}

In financial services, AI solutions are deployed in complex environments that exhibit dynamic, adversarial conditions.  Moreover, AI models increasingly operate in \textit{Systems of AI} with high degrees of interconnectedness and interdependence.  Modern regulatory frameworks, which struggle at even the model level, require new approaches to deal with the unprecedented complexity and unpredictability of such AI systems. As discussed earlier, the GenAI dual-complexity issue causes interlocking of AI regulation with the alignment of the building blocks.

\begin{itemize}
    
\item \textbf{Data Risks:} Generative models amplify data risks due to their high demand for training data and the fact that it is hard to guarantee quality at internet scale \cite{GPT4, GEMINI, Anthropic}. Furthermore, fine-tuning in financial AI models rapidly increases AI safety risks and the likelihood of sensitive data leakage \cite{Duff23}.

\item \textbf{Worsening Hallucinations:} Despite all efforts to the contrary, generative AI hallucinations worsen in more advanced models \cite{MoreHallu}.

\item \textbf{Hidden Discrimination and Persistent Toxicity:} Despite repeated mitigation efforts, financial AI models still show entrenched bias \cite{Gallegos2023} and generate toxic outputs \cite{Ferrer,Wells}. They can also possibly “starve” emerging or legally unprotected groups by systematically withholding resources or opportunities in ways that current oversight tools, tuned to traditional protected classes, fail to spot \cite{Grover95}.

\item \textbf{Increasing Ethics Issues:} Emerging AI ethics problems are often harder to identify and address \cite{Wachter21,Barnes20,Boyd14}.

\item \textbf{Attacks and Safety:} The growing variety and severity of attack vectors threaten the safety of current and future AI deployments \cite{Amodei16,Challen19,Das25}. 

\item \textbf{Lack of Transparency:} Generative AI models are inherently difficult to interpret and their complexity renders traditional explainability methods increasingly ineffective \cite{GeminiLatest}.

\end{itemize}

\subsection{GenAI is a CAS}

In the past decade, transformer models and Kaplan scaling \cite{Kaplan} caused dramatic increases in AI model complexity, until recent scale-down trends \citep{ChengWZZ2017}. This resulted in unprecedented growth in both parameter count and architectural complexity \cite{GPT4Parameters,GEMINI}. Models are now approaching the ten trillion parameter mark, with the number of parameters doubling every year since 2010 \cite{Nature}. 

Generative and agentic AI systems exhibit hallmark features of \textit{complex systems} \cite{ladyman2013}, including \textit{high-dimensional, nonlinear interactions} among internal components such as attention heads and transformer layers, and  \textit{emergent behaviors} that appear as in-context learning and abrupt capability phase transitions. Similarly, they exhibit \textit{open-system characteristics}, as GenAI model behavior is shaped by both  internal architecture and dynamic external inputs. They have \textit{adaptive behaviors} both during training phases through reinforcement learning (e.g. RLHF and RLAIF) as well as during agentic interactions with the environment.

\begin{figure}[t]
\centering
\includegraphics[width=0.6\columnwidth,keepaspectratio]{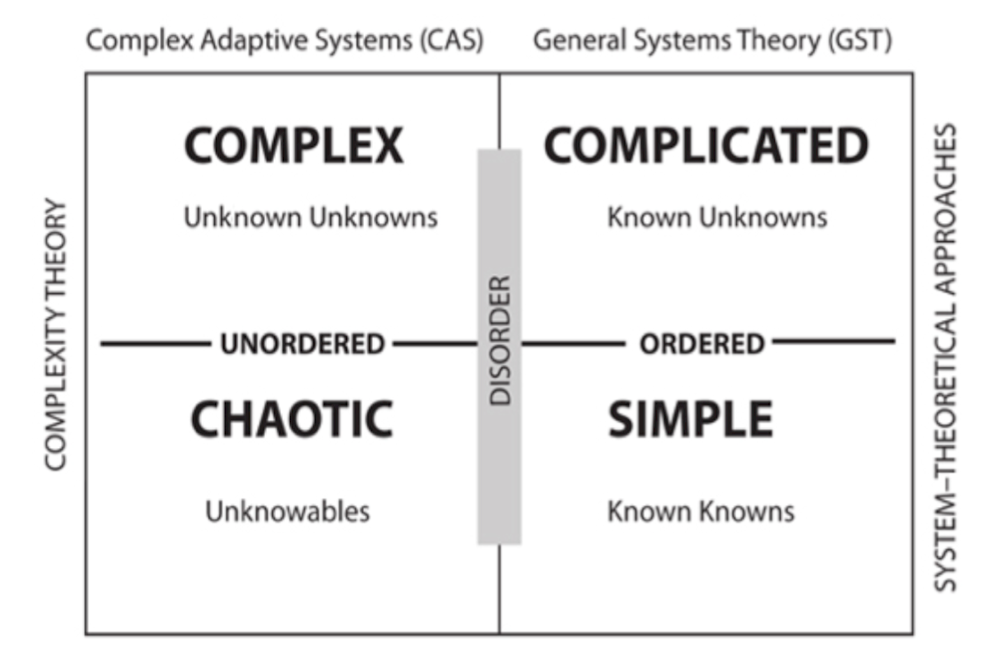}
\setlength{\belowcaptionskip}{-10pt} 
\caption{\small{Categorization of systems: Complex and Complicated system and theory differences \cite{Tele}}}
\label{fig:Fig0}
\end{figure}

Complex systems are fundamentally different from complicated systems (such as high-end computers), because they lack \textit{controllability}, \textit{observability} and mathematical modeling support (\Cref{fig:Fig0}), making them impossible to regulate or govern using traditional model risk management approaches. Complex system regulation (especially high-parameter systems) is an open problem.  Ensuring safe autonomy and effective regulation is becoming a grand challenge for both GenAI and regulatory frameworks.

\subsection{Agentic AI and CAS Risks}

The autonomous objective optimization of Agentic AI systems has caused unpredictable, emergent, and potentially dangerous behaviors, especially in dynamic environments \cite{Zhang2024, Kampouridis2022, Silver16}. Agentic AI exhibits critical vulnerabilities like reward hacking and specification gaming \cite{Pan}, and even leads to criminal behaviors like price collusion \cite{Chica24,Itay25,Banchio23} and market manipulation \cite{Mizuta20,Wellman20,Azzutti22} in financial use cases.  

In recent years, jail-breaking and its variants such as Best-of-N \cite{BON}, prompt injections, backdoor attacks \cite{Saha20} and other AI-Takeover Attacks (AITO) have been reported \cite{Amodei16}. Guardrails and novel techniques have not been able to stop the rapidly growing list of AITO attacks.

Increased AI autonomy raises the risk of deceptive alignment, where a model hides its underlying misalignment, undermining even the most advanced model development and testing processes \cite{Jiu}.

Though related to agentic hacks, emergent unethical behaviors are of special concern. Since the introduction of agency, AI has demonstrated the capacity to independently develop unethical strategies, including deception, lying and others \cite{OBroin07,Trewin22,Wired23,Guardian24,Insurance,Joel17}.  

\begin{figure}[t]
\centering
\includegraphics[width=0.8\columnwidth,keepaspectratio]{fig/Fig2.png}
 \setlength{\belowcaptionskip}{-11pt} 
\caption{\small{High-level framework for a hypothetical AI model: (i) Self-regulation blocks and (ii) Local governance and regulatory blocks. Block names are for illustration purposes.}}
\label{fig:Fig2}
\end{figure}

\section{Proposed Governance Framework}

Previous sections made clear that the feedback loops and emergent behavior of generative and agentic AI are outpacing the linear, one-time validation controls that dominate current financial regulation.  We now pivot from diagnosis to prescription.  Drawing on insights from complex adaptive systems, we introduce a governance architecture that mirrors the decentralization and continual adaptation of the technologies it must oversee.  Oversight functions are wrapped into modular, interoperable \emph{regulatory blocks} that can be composed, upgraded, or retired without disrupting the whole.  These blocks are grouped into four complementary governance layers:

\begin{itemize}
  \item \textbf{Self-Regulation Capabilities} \textit{(Hosted \& Maintained by the Firm)} — tightly integrated local modules that sit beside each model and enforce behavioral, ethical, and performance constraints within milliseconds.
  \item \textbf{Model Governance and In-House Regulation Capabilities} \textit{(Hosted \& Maintained by the Firm)} — firm-managed blocks that guarantee jurisdiction- and application-specific compliance for every model, while supporting in-house regulatory tiers maintained in collaboration with supervisory agencies.
  \item \textbf{External Regulation Capabilities} \textit{(Hosted \& Maintained by Regulators)} — regulator-hosted agents that ingest anonymized telemetry across institutions to detect systemic and network-level risks, including multi-party financial crimes \cite{Aggarwal2006,Cartea2020,Markham1988,Bosch1991,Connor2007,Connor2014}.
  \item \textbf{Independent Audit} \textit{(Hosted \& Maintained by Independent Auditors)} — third-party audit blocks that provide decentralized assurance and are critical for sustaining public trust and long-run AI safety.
\end{itemize}

Next we describe our design strategies, and these layers in turn.  We begin with the principles that guide block construction.

\subsection{Design Strategies}

The proposed governance framework follows a set of design strategies that translate complex adaptive systems (CAS) principles into actionable, modular regulatory mechanisms for generative and agentic AI in finance. Each strategy targets a specific challenge ranging from real-time self-regulation at the model level to systemic risk monitoring across institutions, and aligns with one or more of the framework’s governance layers. In the subsections that follow, we detail eight such strategies, explaining their rationale and operational requirements.

\vspace{4pt}
\begin{Strategy}
Layered Functional Specialization of Regulatory Blocks
\end{Strategy}

Although behavior in complex adaptive systems is emergent, assigning clear functional roles to distinct subsystems is an effective way to retain control.  In our framework this principle translates into three specialized layers: (i) \textit{self-regulation} blocks embedded at the model level, (ii) \textit{governance} blocks that operate at the firm level, and (iii) \textit{regulation} blocks that span the firm and supervisory levels.  By separating these responsibilities, the framework gives risk teams and regulators precise levers for monitoring, intervention, and policy updates, as illustrated in \Cref{fig:Fig2}.

\vspace{4pt}
\begin{Strategy}
Standardization of Components for Complexity
\end{Strategy}

The use of standardization is a promising strategy to deal with complexity in CAS as well. Building libraries of specialized regulatory blocks tailored to each model’s application and jurisdiction, such as Equal Credit Opportunity Act (ECOA) blocks \cite{ECOA} for mortgage or credit models, can help with AI regulation. These blocks (which can include AI agents) enable continuous compliance monitoring, real-time feedback, and enforcement, promoting a consistent regulatory framework across applications. Governance versions of the blocks are developed by the firms or vendors, and regulatory versions by the regulators. 

 It's important to note that regulatory blocks cover the full agency spectrum from no-agency rule-based, hard-coded blocks, to limited agency and agentic blocks. Furthermore, agentic blocks cover a range of agency and can be configurable during run-time. Despite the flattened illustration in \Cref{fig:Fig2} interactions among blocks can be quite complex. However, these design specifications are expected to be optimized for each model, pursuant to regulatory needs and system goals such as safety and reliability.

Standard regulatory block libraries enable coverage across multiple regulation types: 

\textit{(i) Data Regulation Blocks:} Spectrum of regulatory blocks and agents for GDPR \cite{GDPR}, CCPA \cite{CCPA}, RFPA \cite{RFPA}, etc.

\textit{(ii) AI Regulation Blocks:} Blocks and agents for laws like the EU AI Act \cite{EUAI} and Singapore AI regulations \cite{SingaporeAI}.

\textit{(iii) Financial Regulation Blocks:} Application- and jurisdiction-specific blocks and agents for ECOA \cite{ECOA}, HDMA \cite{HDMA}, BSA \cite{BSA}, and regulatory guidelines from agencies like the Federal Reserve and OCC.

\textit{(iv) Local MRM Agents:} Internal agents for firm-specific policies and model risk management rules. Standardization of blocks does not imply uniform blocks with similar characteristics. Regulatory blocks may have highly varying characteristics in terms of size, complexity, capabilities, architecture, autonomy, and interactivity.  While some blocks may be simple rule-based systems, others may be complex agentic solutions with internal hierarchies and numerous components with specialized roles.

\vspace{4pt}
\begin{Strategy}
Modular Design for Change Management, Updates and Customizations
\end{Strategy}

Modularity is an effective design strategy for adapting to dynamic environments, enabling faster deployment, lower costs, and faster entry into new markets or product lines.

As global AI and financial regulations expand, models must operate across diverse geographies and evolving regulatory landscapes. Capturing regional and temporal variations is complex, and rising AI development costs make region-specific models increasingly impractical. (a) Modular regulatory components enable updates and replacements without adding complexity to the core AI model, supporting third-party integration while reducing compliance and safety risks. (b) They can be used to train and update the models through reinforcement learning.  This can be used for: 

\textit{(i) Regulatory Changes:} Agents covering both financial and AI regulations as well as changes to regulations;
\textit{(ii) Jurisdictions and Geographical Adoption:} Agents customized for regions such as the U.S., European Union, China, etc.;
\textit{(iii) Firm-wide Policy Changes:} Agents enforcing organization-specific policies and regulatory goals across all models, including third-party vendor models.

\vspace{4pt}
\begin{Strategy}
Adaptive System Components Managed by Local Control for Dynamic Environments
\end{Strategy}

\textit{(i) System-wide Agency Adaption:}  System-level adaptiveness to autonomy allows adjusting for different environments and time periods.  In high-risk, adversarial, client-facing, or dynamic environments, autonomy is constrained to enforce human-in-the-loop approvals, rule-based decisions, or supervisory control. Conversely, in low-risk, controlled in-house settings, higher levels of autonomy are permitted. The agentic adaption mechanisms include but not limited to: \textit{Decision-Making Autonomy} based on task-specific risks, \textit{Goal Setting Autonomy} for predefined goal specification constraints, and \textit{Response to Novel Situations} allowing human intervention during abrupt changes.

\textit{(ii) Block-level Custom Agency Adaption:} Adapting degrees of agency at the block level is essential to ensuring enhanced safety and trustworthiness of the overall system. This translates to a new system design and dynamic optimization problem to potentially improve overall characteristics.

\textit{(iii) Temporal Agency Adaption:} Dynamic agency adjustment is required to address emerging safety and compliance concerns. As AI systems evolve, temporal agency control ensures continuous adaptation of the framework, optimizing each generation based on the capabilities of individual components and the evolving application context.

\textit{(iii) Human-in-the-Loop Adaption:}  Certain regulations, such as fair lending laws and KYC-AML rules, mandate human review for all or high-risk cases (e.g., U.S. ECOA \cite{ECOA}, BSA \cite{BSA}, PATRIOT Act \cite{PATRIOTAct}, AMLD \cite{AMLD}). Given the inherent safety risks of autonomous AI, human oversight and control over autonomy remains essential. 

\vspace{4pt}
\begin{Strategy}
Decentralized Architecture to Improve System Safety and Reliability
\end{Strategy}

As discussed earlier, all AI models have the underlying risk of AI Take-over attacks (AITO) and can be used to facilitate criminal and unethical acts \cite{BON}. This risk dramatically increases alongside model autonomy, necessitating the decentralization of regulatory architecture.  This can ensure the safety and robustness of the regulatory system and allow effective oversight, mitigation, and auditing.  Integrating independent monitoring and regulatory hierarchies reduces risks linked to centralized hierarchical control and prevents vulnerabilities from the takeover of a limited set of regulatory blocks.  Financial services heavily depend on a limited set of third-party and vendor models, making them particularly vulnerable to attacks on financial and regulator models. 

\begin{figure}
\centering
\includegraphics[keepaspectratio,width=0.8\columnwidth]{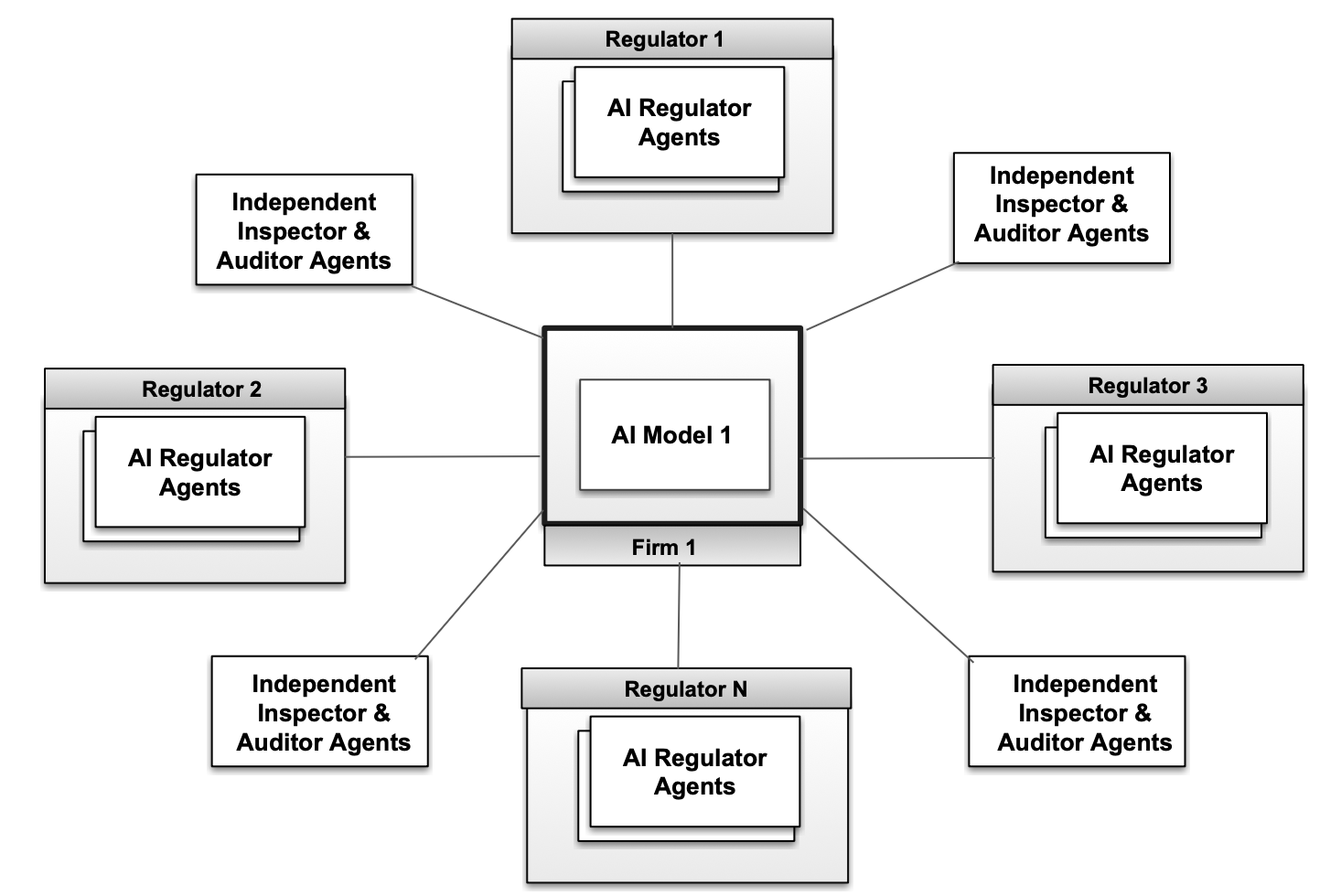}
 \setlength{\belowcaptionskip}{-6pt} 
\caption{\small{Decentralization of financial AI regulation through numerous regulators and independent audit and regulatory blocks and agents}}
\label{fig:Fig4}
\end{figure}

As shown in \Cref{fig:Fig4}, independent regulatory blocks and agents including independent auditors and inspectors \cite{Brundage20,Mokander21} improve safety, which is especially vital for agentic AI given the increasing threat of takeover attacks (e.g. back-door attacks, hidden triggers \cite{Lin20, Saha20, Guo22}), emergent misaligned behaviors, and deceptive alignment challenges \cite{Bogen18}.

\vspace{4pt}
\begin{Strategy}
Diversity of System Components and Agency for Robustness
\end{Strategy}

Structural heterogeneity is a well-known strategy that works in both general system design and complex adaptive systems theory. In the context of regulatory framework this translates to various aspects of the design:

\textit{(i) Heterogeneity of Autonomy:} Heterogeneous autonomy levels (from rule-based blocks to agentic blocks with varying degrees of autonomy) is an important feature for robustness. Autonomy can be tailored for block level functions based on their risk levels, becoming a design parameter to optimize overall system robustness, controllability and safety.

\textit{(ii) Regulatory Heterogeneity:} A single, centralized regulatory layer creates a point of failure and amplifies correlated vulnerabilities. In contrast, the framework disperses oversight across multiple, independently designed regulatory and audit blocks, each with its own control logic and data pathway. This structural diversity hardens the system against coordinated attacks and unexpected faults.

\textit{(iii) Block-level Heterogeneity:} Heterogeneity of the architectural and behavioral characteristics as well as vendors at the  block level also provides a mechanism to hedge against system-level takeover attacks.  

\vspace{4pt}
\begin{Strategy}
Redundancy for Reliability, Fault-Tolerance and Safety
\end{Strategy}
Incorporating redundant building blocks is an effective strategy for complex systems in dealing with the underlying reliability, robustness and safety issues. Not only does this strategy ensure fault-tolerance when one or more components of the system fail or are taken-over during an attack, it provides the ability to isolate and contain faulty or compromised blocks, regions or subsystems.

As the reliability of the building blocks (from the core models to the regulatory framework components) remain an issue, having optimized redundancy levels at different granularities and functional capabilities is highly important to (i)\textit{Block-level Checks} that cross validate the individual block outputs, \textit{Anomaly Detection and Reporting} to report abnormal behavior, and \textit{Safety Issues} for detecting potential safety and security vulnerabilities and issues as they emerge. Additionally, redundancy also provides better performance in dynamic adversarial environments during functional failures as well.

\subsection{Self-Regulation Within Agents}

Self-regulation modules consist of embedded monitors, ethics filters and safety guards that are wired directly into the runtime of each generative or agentic model.  Because they operate in the same execution context, they can intervene within milliseconds, a capability that is indispensable when trying to govern the highly unpredictable behavior of complex adaptive systems such as modern GenAI models.  The challenge is amplified by the growing mix of in-house and vendor models whose trustworthiness, safety and reliability vary widely, creating a heterogeneous risk surface that traditional, centralized controls struggle to manage \cite{Hentzen2022,Christensen2021,Fernandez2019,Ferreira21}.  By imposing a common set of local controls, self-regulation blocks provide a uniform policy.

Each block delivers four tightly coupled functions.  First, it performs core monitoring and regulation, supervising performance drift, data-quality anomalies and environmental changes, and throttling or quarantining the model when predefined bounds are breached.  Second, it conducts ethics self-regulation, continuously inspecting inputs and outputs for fairness, bias, explainability and broader ethical compliance.  Third, it enforces safety and security self-regulation by detecting prompt-injection, data-exfiltration and other unsafe behaviors, then invoking automated or human-in-the-loop responses.  Finally, every intervention is written to an immutable audit log so that internal risk teams and external regulators can reconstruct events, assess control effectiveness and refine policies over time.  Collectively, these capabilities make each model the first line of defense, aligning day-to-day operation with enterprise policy and statutory requirements long before risks can propagate to higher tiers of governance.

\subsection{Firm-Level Governance Modules}

Firm-level governance modules sit between local self-regulation blocks and external regulators, giving the institution an enterprise view of model behavior, risk posture, and compliance status.  They ingest real-time telemetry from thousands of self-regulation modules such as latency metrics, ethics flags, security alerts, then fuse those signals with business context such as position limits, customer segmentation, and liquidity exposure.  The resulting dashboards allow risk and compliance officers to spot correlated failures, trigger circuit breakers, or throttle model access when aggregate indicators breach predefined tolerances.  Because these modules are hosted and maintained by the firm, they can enforce jurisdiction-specific requirements (for example, Europe’s AI Act or U.S. fair-lending rules) while still aligning with global policy frameworks such as SR 11-7 and Basel Principles \cite{SR11-7}.  They also provide a single point of attestation for board governance committees, converting raw telemetry into evidence packages suitable for annual model-risk review.

Architecturally, each firm-level module exposes standard APIs, REST for data pulls, gRPC or message queues for event pushes, so that new models or vendor services can register without custom wiring.  A rule engine or policy microservice translates high-level mandates (for example, maximum daily loss or bias thresholds) into machine-readable checks that propagate back down to self-regulation blocks.  The same interface allows external regulators and independent auditors to query anonymized, aggregated metrics without breaching data-protection walls.  By decoupling policy logic from individual models, the firm can roll out updated controls by hot-swapping container images or feature flags rather than redeploying every model.  This design supports rapid iteration as legislative guidance evolves, while maintaining a continuous audit trail that links each control decision to the policy version in force at the time of execution.

\subsection{External Regulatory \& Audit Modules}

External modules are hosted by supervisory agencies or accredited third-party auditors and operate at the sector or market level. They collect anonymized, aggregated telemetry streams and event logs from multiple firms, enrich those data with public market feeds, and apply anomaly-detection models to uncover collusive behavior, emerging concentration risk, or coordinated cyber threats that no single institution could observe in isolation. Each module publishes versioned policy APIs through which regulators can issue compulsory control updates, such as new bias thresholds or liquidity stress tests, that downstream governance layers must enforce. Cryptographically signed attestations and tamper-evident ledgers ensure that audit evidence remains verifiable while preserving firm confidentiality through differential privacy or secure multiparty computation techniques. By providing a neutral vantage point, these modules close the information gap between micro-prudential oversight and macro-prudential stability, and they furnish an independent assurance layer that strengthens public trust in the safety of AI-driven financial services.

\section{Case Study: Emergent Unethical Behavior}

To illustrate the potential of our modular regulatory framework, in this section we explore a specific set of problems in the literature, discuss potential solution components, and show how they might fit into the proposed framework.

\subsection{Agentic Misbehavior in Financial Markets}

Multiple lines of research have examined the disruptive behavior potential of intelligent trading agents in a financial market system.  Spoofing, or the manipulation of market prices by placement of orders not intended for transaction, has been especially studied.  Using the Market-Sim simulation \cite{wah2017welfare}, a research group found that a generative order-placement model could not only learn to spoof its market, but to hide such behavior as legitimate market making \cite{wang2020market}.  Other investigators, using the ABIDES simulation \cite{byrd2020abides}, found that a reinforcement learning (RL) based trader tasked with simple profit maximization, could discover and employ spoofing as a strategy so long as it could place and cancel limit orders.  Scholars have also investigated the effect of spoofing on market efficiency and price discovery \cite{liu2022spoofing} and of liquidity on spoofing \cite{gu2024effect}.  Taken together, these studies have shown that agentic spoofing, inadvertent or otherwise, arises often and with a disruptive impact on other market participants, making markets less efficient, hampering price discovery, and increasing volatility.

\subsection{Remediation Efforts}

There have also been preliminary efforts to redress the discovery of spoofing as a profit-maximizing strategy by intelligent trading agents.  Using the generative model discussed above \cite{wang2020market}, the investigators found that it was effectively countered by an adversarial discriminator tasked with spoofing detection.  To evade detection, the generative model was forced to eventually become an honest market maker.  For the RL based trader \cite{byrd2022learning}, the authors drew techniques from the field of Normative RL, applying reward shaping, policy shaping, and action reranking to the agent training process.  These methods relied on an internal feedback mechanism, trained in simulation with known spoofing strategies, which discouraged selection of actions that produced a trajectory similar to spoofing, and were effective in significantly reducing the appearance of such behaviors.

\begin{figure}
\centering
\includegraphics[keepaspectratio,width=0.8\columnwidth]{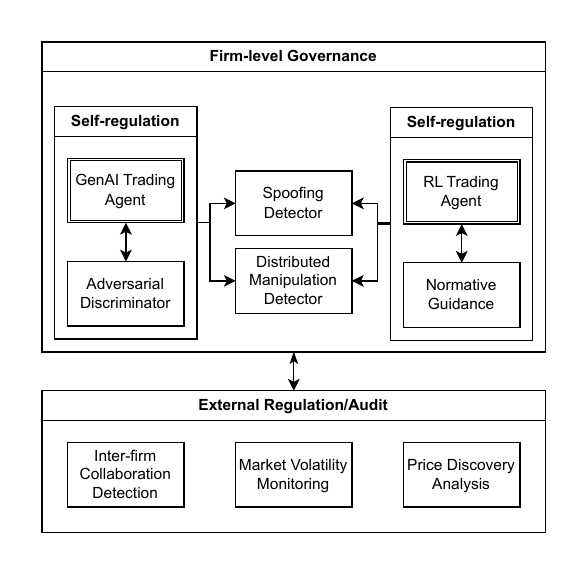}
 \setlength{\abovecaptionskip}{-6pt}
 \setlength{\belowcaptionskip}{-6pt} 
\caption{\small{Example of GenAI spoofing risk mitigation}}
\label{fig:CaseStudySpoofing}
\end{figure}

\subsection{Framework Incorporation}

The results of these studies suggest components for inclusion into the proposed modular governance framework, illustrated in Figure \ref{fig:CaseStudySpoofing}.  Both the adversarial discriminator and the RL normative guidance component would make excellent \emph{self-regulation} blocks, residing adjacent to their individual models. The feedback from these blocks could directly influence learned model behavior, improving model risk management.  Generalized versions of the discriminator and the spoofing detector are candidates for \emph{firm-level governance} blocks.  Rather than providing direct model feedback, these could combine firm-wide order flow data to guard against distributed spoofing collusion among agents, which would be undetectable at a lower level.  These could provide feedback to the firm before deployment during a formal model validation exercise or after deployment based on live market activity.  Further variations of these components trained to recognize not only aggregated spoofing behavior signatures across firms, but also the aforementioned impacts on volatility, efficiency, and price discovery, might comprise \emph{external regulatory} blocks, logging detected patterns and raising an audit flag for human investigation if the target behaviors are ongoing.  These and other regulatory blocks can provide feedback independently or be wrapped into ensembles using meta-learning or mixture-of-experts approaches for more comprehensive surveillance and guidance.

\subsection{Framework Benefit}

This simple example discusses only one small area (emergent spoofing behavior in AI trading agents) of the vast governance space about to be challenged by rapid AI deployment.  Given the scope of the issue and the pace of change, neither a traditional model nor a monolithic technical model of governance is likely to succeed.  The modular framework is adaptable, responsive, and can grow and change alongside technical developments in the AI Finance space.

\section{Further Capabilities}

\label{sec:opportunity}

Here we outline additional capabilities of our modular framework that extend traditional governance: supporting consistent oversight, real-time monitoring, safety and security, end-to-end lifecycle management, and rigorous testing.

{\bf Consistent Regulatory Oversight}: AI governance remains fragmented, with inconsistent standards, reactive scrutiny, and diverse implementations across firms. Standardizing tools and processes is critical to improving safety and compliance.

Our framework consolidates oversight through modular blocks that integrate third-party tools and firm-specific governance systems. This enables scalable, jurisdiction-aligned governance across diverse models and deployments.

{\bf Real-Time Monitoring \& Regulation}: Agentic AI requires continuous oversight due to adaptive, autonomous behavior. Our framework supports:

\begin{itemize}
\item \textit{Core Model Monitoring:} Tracks performance metrics often missing in existing tools \cite{Abiyoke}.

\item \textit{Automated Reporting:} Generates summaries and metrics for regulators and compliance.

item \textit{Agentic Trust Monitoring:} Evaluates metrics such as task success \cite{Kwa}, error awareness \cite{Tamer}, stress robustness \cite{Lakshmi}, consistency \cite{Galileo}, and harmful outputs \cite{CMARIX}.
\end{itemize}

The framework supports independent safety and security agents:

\begin{itemize} 

\item Action monitoring for misuse;

\item Model safety for interaction threats;

\item Security oversight to detect attacks.
\end{itemize}

{\bf Lifecycle Governance}: Lifecycle governance, is now a regulatory necessity \cite{ChinaAI}. Our framework supports oversight from development through deployment—covering training data, specifications, testing, RLHF risks, and profiling—via audit-ready reporting blocks aligned with policy requirements.

Rigorous testing is essential to uncover risks in complex models \cite{Gao,Hannon2023}. Our framework enables:

\begin{itemize}

\item Adversarial testing engines;

\item Expanded and multi-criteria evaluations;

\item Real-time stress testing akin to CCAR \cite{CCAR};

\item Behavioral profiling and standardized AI test coverage.
\end{itemize}

{\bf Response to Rapid Advancement}: Rabid model advancement \cite{Llama4,GPT4,Anthropic,GeminiLatest,mistral,DeepSeek} challenges traditional oversight. Financial firms often deploy high-risk models without proper controls. Our layered regulatory architecture supports adaptive governance while minimizing risks tied to evolving model behavior and deployment speed.

\section{Conclusions \& Outlook}
\label{sec:conclusion}

AI systems are now integral to financial services. As generative and agentic AI adoption grows, effective model governance is critical to ensure robustness and regulatory compliance. Model complexity and regulatory requirements have rendered traditional governance processes ineffective, intensifying concerns about reliability and trustworthiness. 

\vspace{2pt}

As a \textit{high-parameter complex system}, GenAI is upending traditional model risk management and regulatory frameworks. It leads to an open problem for AI regulation, which is strongly intertwined with other underlying challenges like alignment. Despite limited and inadequate regulatory oversight, generative AI solutions are being deployed at increasing rates, amplifying risks that require urgent and proactive mitigation.

\vspace{2pt}

This paper surveys the numerous challenges in regulating generative and agentic AI. It presents complex adaptive system strategies and a high-level hybrid framework for their integration as an agentic regulator. The framework aims to tackle the growing gaps in current model risk management and regulatory approaches while addressing complexity, agility, reliability and safety challenges as primary goals.



\bibliography{bib}
\bibliographystyle{ACM-Reference-Format}


\end{document}